\documentclass[aps,prb,twocolumn,superscriptaddress,longbibliography]{revtex4-2}

\usepackage{listings}
\usepackage{xcolor}
\usepackage[utf8]{inputenc}
\usepackage{placeins}

\usepackage{physics}
\usepackage{comment}
\usepackage{amssymb}
\usepackage{booktabs}
\usepackage{multirow}
\usepackage{float}
\usepackage{ulem}

\usepackage[colorlinks=true,citecolor=blue]{hyperref} 
\usepackage{graphicx}
\usepackage{amsmath}
\usepackage{dcolumn}
\usepackage{bm}

\newcommand{\newc}{\newcommand}
\newc{\kt}{\rangle}
\newc{\br}{\langle}
\newc{\pr}{\prime}
\newc{\longra}{\longrightarrow}
\newc{\ot}{\otimes}
\newc{\rarrow}{\rightarrow}
\newc{\h}{\hat}
\newc{\bom}{\boldmath}
\newc{\btd}{\bigtriangledown}
\newc{\al}{\alpha}
\newc{\be}{\beta}
\newc{\ld}{\lambda}
\newc{\sg}{\sigma}
\newc{\p}{\psi}
\newc{\eps}{\epsilon}
\newc{\om}{\omega}
\newc{\mb}{\mbox}
\newc{\tm}{\times}
\newc{\hu}{\hat{u}}
\newc{\hv}{\hat{v}}
\newc{\ii}{\dot{\iota}}
\newc{\cf}{{\cal{F}}}
\newc{\md}{\mbox{D}}
\newc{\RNum}[1]{\uppercase\expandafter{\romannumeral #1\relax}}

\begin{document}

\title{Machine-learning-enabled characterization of individual ring resonators in integrated photonic lattices.} 

\author{Elizabeth Louis Pereira}
\affiliation{Department of Applied Physics, Aalto University, 02150 Espoo, Finland}

\author{Amin Hashemi}
\affiliation{CREOL, The College of Optics and Photonics, University of Central Florida, Orlando, FL 32816, USA}

\author{Faluke Aikebaier}
\affiliation{Department of Applied Physics, Aalto University, 02150 Espoo, Finland}
\affiliation{Computational Physics Laboratory, Physics Unit, Faculty of Engineering and Natural Sciences, Tampere University, FI-33014 Tampere, Finland}
\affiliation{Helsinki Institute of Physics P.O. Box 64, FI-00014, Finland}

\author{Hongwei Li}
\affiliation{Nokia Bell Labs, 21 JJ Thomson Avenue, Cambridge, CB3 0FA, UK}

\author{Jose L. Lado}
\affiliation{Department of Applied Physics, Aalto University, 02150 Espoo, Finland}

\author{Andrea Blanco-Redondo}
\affiliation{CREOL, The College of Optics and Photonics, University of Central Florida, Orlando, FL 32816, USA}

\date{\today}
\maketitle

\textbf{
Accurately determining the underlying physical parameters of individual elements in integrated photonics is increasingly difficult as device architectures become more complex. Inferring these parameters directly from spectral measurements of the system as a whole provides a practical alternative to traditional calibration, allowing characterization of photonic systems without relying on detailed device-specific models. Here, we introduce a supervised machine-learning strategy to
learn the onsite losses and resonant frequency shifts of each individual ring in an array of coupled ring resonators from measured spectral power distributions of the whole array. The neural network infers these parameters with high accuracy across multiple experimental configurations. Our methodology
provides a scalable and non-invasive method for extracting intrinsic parameters in coupled photonic platforms,
paving the way for future development of automated calibration and control methods.}

Photonic integrated circuits (PICs) combine multiple optical components on a single chip, enabling fast, compact, and scalable light control. Modern PICs integrate thousands of components and represent a keystone in communications\cite{Chovan2018,Terrasanta2025}, sensing\cite{Milvich2021}, quantum technologies\cite{Giordani2023, Dutta2025}, and optical accelerators\cite{Ning2024}. Many PICs functionalities rely on coupled ring resonators such as  microwave RF filters\cite{Liu2023,Wei2025}, tunable lasers\cite{Hulme2013}, high-speed modulators\cite{Hu2012,Cheng2023}, and integrated frequency combs\cite{Brasch2016,Kippenberg2018,Yang2024}. 
In such complex coupled systems, extracting the individual parameters of each ring resonator, such as their intrinsic losses and resonance frequencies, is essential for achieving high-performance of the system as a whole. Although multiple ingenious methods have been proposed for extracting these key parameters based on precise optical characterization \cite{McKinnon2009,Li2013,Fedorov2025,Umetaliev2025}, the fast and automated extraction of internal parameters remains a challenge with traditional fitting techniques as the complexity of the circuit increases.

Machine-learning (ML) approaches are increasingly being explored in the characterization and control of integrated photonic systems because they offer a way to model device behavior directly from experimental data rather than from idealized coupled-mode or transfer-matrix descriptions\cite{Fu2023}. Prior work has shown that learning-based models can handle nonlinear spectral distortions, fabrication variability, and cross-talk effects that are difficult to capture with traditional analytical tools\cite{Hunter2025}. More broadly, ML-assisted design and parameter extraction has been applied across quantum\cite{Torlai2018,Carrasquilla2019,PhysRevA.105.023302,PhysRevResearch.1.033092,PhysRevResearch.3.023246,Gebhart2023,PhysRevLett.122.020504,PhysRevLett.127.140502,Nandy2024,Huang2022,f58h-zxs3}, electronic\cite{PhysRevApplied.20.044081,vanEsbroeck2020,2024arXiv240504596V,v11m-dbhm,PhysRevX.14.011001,PhysRevB.110.075402,Moon2020,13c4-p4fq}, and photonic systems\cite{Wiecha2021,Khan2022,book,Zhou2024,Wang2024,Almasi2024}, where it enables fast inference of Hamiltonian parameters from measured responses\cite{Khosravian2024,PhysRevApplied.23.054077,Koch2025,PhysRevResearch.4.033223}. However, a method for the characterization of the physical parameters of each element in large coupled systems has remained elusive.  

Here, we establish a scalable ML methodology
to infer intrinsic losses and resonance shifts in coupled ring-resonator arrays. We implement this on a programmable integrated photonic platform\cite{capmany2020programmable, On2024,amin2024, Love2025}.
In contrast with
previous ML models targeting parameter extraction in single devices\cite{https://doi.org/10.48550/arxiv.2506.17999,Fyrillas2024,Zhang2023,FerreiradeLima2022}, our approach captures the collective response of an entire array, enabling simultaneous and non-invasive reconstruction of multiple system parameters at once.
Our algorithm is directly trained on experimental data from a programmable photonic chip, removing the need for
synthetic theory data to extract the physical parameters. To validate our method, we demonstrate a second algorithm
that enables the reconstruction of the experimental spectra from the theoretically simulated spectra, thus serving as a measure of the quality of our method by comparing the reconstructed and measured spectra.
This strategy to extract and assess the Hamiltonian of a programmable photonic system provides a scalable method for the automated characterization and control of complex photonic circuits.

\begin{figure}[]
\includegraphics[width=\linewidth]{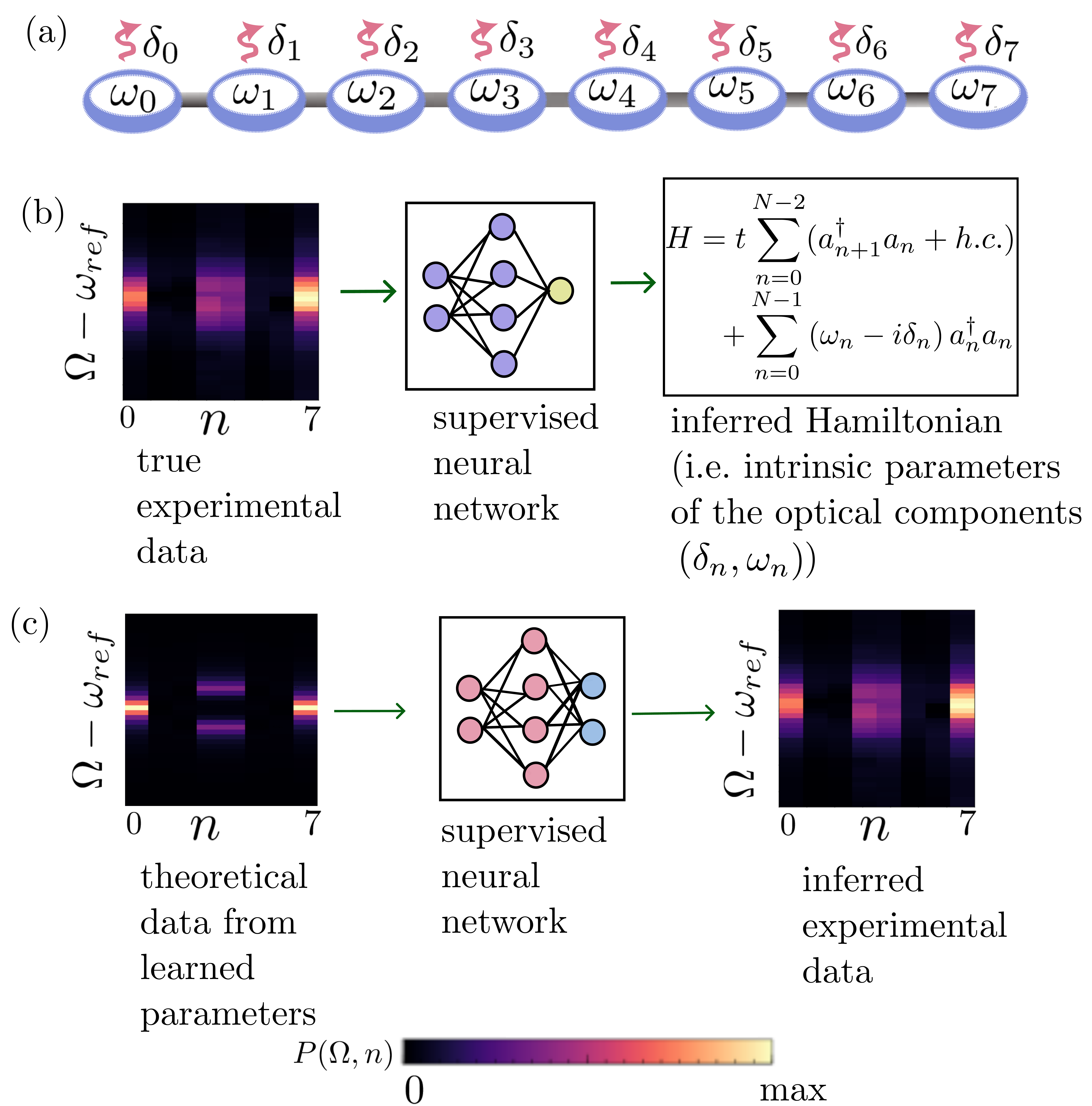}
\caption{(a) Schematic of a system of $8$ coupled ring resonators with onsite losses $\delta_n$ and resonant frequencies $\omega_n$ for each $n$-th ring. 
(b) Schematic of the machine-learning approach.
Experimental spectral power $P(\Omega,n)$ (as shown in color) from the programmable photonic platform \cite{amin2024} is used to train and test the model; the network learns to infer the intrinsic parameters like losses and resonant frequencies from measured data of an array of $8$ optical components. (c) Schematic of the inverse problem where we 
map simulated theory data to the expected experimental data,
and compare it with the true experimental data. Here, $\Omega$ is the input frequency, $n$ is the ring index, and $\omega_{ref}$ is the reference frequency.
}
\label{panel:1}
\end{figure}

\section{Model for non-Hermitian coupled ring resonators}

The PIC we focus our experiments on realizes a set of $8$ coupled ring resonators, as shown in Fig.~\ref{panel:1}(a), and is implemented on a programmable platform \cite{capmany2020programmable,On2024,amin2024}.
The propagation of light can be described by the 
following Hamiltonian, 
\begin{multline}
     H = t\sum_{n=0}^{N-2}\left(a_{n+1}^{\dagger}a_{n}+h.c.\right) + \sum_{n=0}^{N-1}\left(\omega_{n}-i\delta_{n}\right)a_{n}^{\dagger}a_{n} 
    \label{hamiltonian} 
\end{multline}
where $\delta_n$ and $\omega_n$ are the local onsite loss and resonant frequency, respectively, of the site $n$,
and $t$ is the coupling strength, that
we take uniform across the chain for concreteness. 
The operators $a_{n}^{\dagger}$ and $a_{n}$ are the photon
creation and annihilation operators on site $n$.
For our experiments, we fix $N=8$ and $t=5.34$ GHz.

The Hamiltonian above models a one-dimensional chain of eight uniformly coupled ring resonators. The real part of the diagonal elements of the Hamiltonian describe the onsite resonant frequencies of individual rings $\omega_n$, which may vary from site to site due to fabrication disorder or intentional detuning\cite{Yariv1999}. Their imaginary components $\delta_n$ represent both intrinsic loss, primarily due to radiation leakage and scattering within the resonator\cite{Borselli2007,Wang2021}, and coupling loss from the rings to the input/output ports\cite{ElGanainy2018,ourprr,amin2024}. These losses make the system effectively non-Hermitian and its control
enables engineering tunable dissipation\cite{Miri2019}.
The off-diagonal terms of the Hamiltonian encode evanescent coupling between adjacent rings, enabling photon tunneling across the array. This interaction gives rise to extended collective modes whose properties depend on the balance between coupling and loss. The Hamiltonian, therefore, captures the essential physics of light propagation, interference, and attenuation in a lossy resonator lattice, providing a minimal framework to study how non-Hermitian effects reshape spectral and spatial mode structure.

Next, we analyze the characteristic spectral response of the system. The output power of the $n$-th ring, $P(\Omega,n)$, is obtained from the
Green's function $G_n(\Omega)$ as\cite{amin2024}
\begin{equation}
    P(\Omega,n) \propto |G_{n}(\Omega)|^{2}=\left|\sum_{j} \frac{\langle n|\psi_{j}^{R}\rangle \langle \psi_{j}^{L}|n\rangle }{\Omega-\epsilon_{j}}\right|^{2},
    \label{green}
\end{equation}
where $\Omega$ is the frequency, $\bra{\psi_{j}^{L}}$ and $\ket{\psi_{j}^{R}}$ are the left and right biorthogonal eigenvectors of the Hamiltonian $H$ and $\epsilon_j$ is the complex eigenvalue.
This formulation links each spectral feature to specific eigenmode contributions, allowing the measured power distribution to be interpreted directly in terms of the underlying non-Hermitian modes and revealing how losses and resonance shifts shape the observed response.

\section{Learning Hamiltonian parameters from experimental data}
\begin{figure}[]
\includegraphics[width=\linewidth]{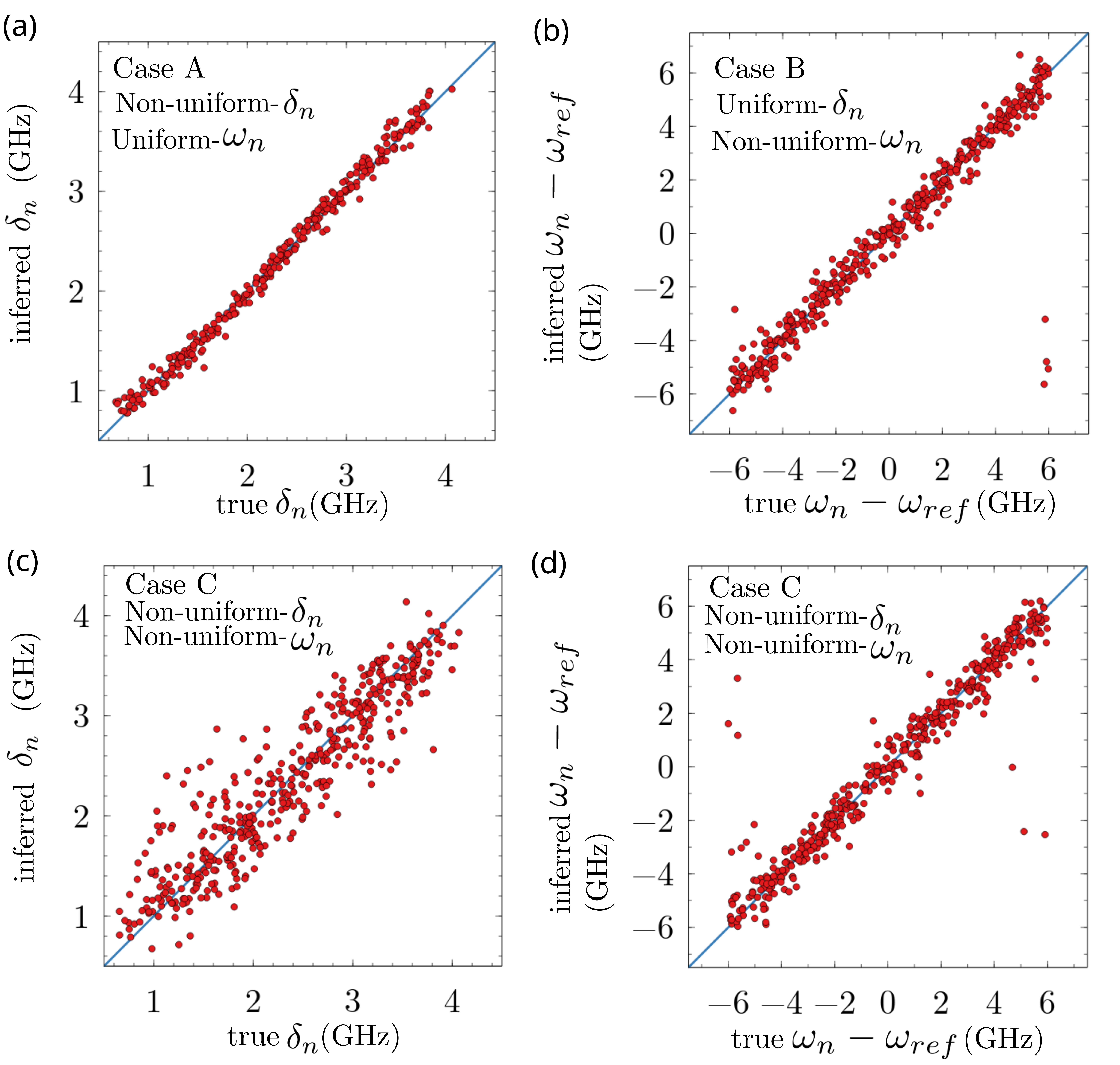}
\caption{(a) Inferred onsite losses $\delta_n$ (red dots) versus true values (blue line) for \textit{case A}, where all rings share the same resonant frequency $\omega_n$. (b) Inferred resonant frequencies versus true values for \textit{case B}, where all rings share the same loss. (c) and (d) Inferred losses and resonant frequencies, respectively, versus true values, for case C, where both loss and resonant frequencies vary across rings.}
\label{panel:2}
\end{figure}

 Our model uses a neural-network algorithm \cite{Schmidhuber2015} to provide direct mapping between experimentally measured spectral densities and intrinsic individual resonator parameters, as shown in Fig.~\ref{panel:1}(b). For our experiments, we configured the system in three different regimes: \textit{Case A} implements an 8-ring resonator array with uniform resonant frequency for all rings $\omega_n=\omega_{\mathrm{ref}}=193,451$ GHz and randomly varying losses for each ring $\delta_{n}$. \textit{Case B} implements an array with uniform loss $\delta_n=\delta=0.49$GHz and varying resonant frequencies for each ring. Lastly, \textit{case C} implements an array with varying losses and resonant frequencies across resonators. In each of these regimes, we generated random configurations between $1600$ and $2400$ and measured the spectral response of the system $P(\Omega,n)$ for each of them. These experimentally measured responses were used to train a fully connected neural network, which in turn performed the physical parameter inference tasks. Fig.~\ref{panel:2} shows the results of these inference tasks in the three different regimes.
Each model uses 40 features as input (the frequency grid $\Omega$ is of size $40$) and was optimized using mean-squared-error loss with ReLU activations\cite{10.5555/3104322.3104425,pmlr-v15-glorot11a}, Glorot initialization, and the Adam optimizer\cite{Kingma2014AdamAM,j.2018on}. The code used in this work and the data for the figures is publicly available on Zenodo \cite{pereira_picml_2026,elizen}.

We characterize the quality of these inference tasks by using the fidelity metric\cite{Khosravian2024,PhysRevApplied.23.054077,Koch2025} defined as
\begin{equation}
\begin{split}
&{\cal{F}}\left(\lambda^{\mathrm{NN}},\lambda^{\mathrm{exp}}\right)\\&=\frac{\left|\langle\lambda^{\mathrm{NN}}\lambda^{\mathrm{exp}}\rangle\right|-\langle\lambda^{\mathrm{NN}}\rangle\langle\lambda^{\mathrm{exp}}\rangle}{\sqrt{\left(\langle(\lambda^{\mathrm{NN}})^{2}\rangle-\langle\lambda^{\mathrm{NN}}\rangle^{2}\right)\left(\langle(\lambda^{\mathrm{exp}})^{2}\rangle-\langle\lambda^{\mathrm{exp}}\rangle^{2}\right)}}\end{split},
\label{fidelity}
\end{equation}
where $\lambda^{\mathrm{NN}}$ is the value extracted by the neural-network
of the Hamiltonian parameter $\lambda$ and $\lambda^{\mathrm{exp}}$ its experimental value.
In our case, $\lambda=\delta_n,\omega_n$ takes the intrinsic loss or the resonance frequency.

As seen in Fig.~\ref{panel:2}(a) and (b), the neural network can reliably infer the onsite losses and resonant frequency shifts, respectively, directly from the measured spectra when one of the parameters (resonant frequency shift or onsite loss, respectively) is fixed. The calculated fidelity in these two cases are ${\cal{F}}=0.99$ and ${\cal{F}}=0.94$, respectively.

Case C, shown in Fig.~\ref{panel:2}(c) and (d), highlights a physical asymmetry in the inverse problem. Non-uniform variations in the onsite losses do not significantly affect the quality of inference of the resonant frequency, as shown in Fig.~\ref{panel:2}(d), which yields ${\cal{F}}=0.95$. However, non-uniform variations of the resonant frequencies have a negative impact on the quality of the inference of the onsite loss, as shown in Fig.~\ref{panel:2}(c). In the latter case, the fidelity of the onsite loss prediction decreases to $0.92$. This asymmetry can be explained by the fact that the frequency detuning affects the interferometric mesh globally, altering the local spectral contrast used to infer losses. However, losses remain more spatially localized and have weaker influence on adjacent resonance conditions.
Our results show that the spectral response of the coupled system contains sufficient information for accurate parameter inference without additional calibration or probing. The results also highlight the physical interdependence between parameters, with the neural network inferring losses and resonance frequencies with the same dependencies that exist in the actual photonic platform.

Our machine-learning approach uses the experimentally measured spectrum of an individual ring resonator within a coupled system to infer its intrinsic parameters. Since the local spectrum already contains implicit information about neighboring resonators through inter-resonator coupling, the same inference strategy can, in principle, be generalized to larger and more complex photonic integrated circuits. In the present work, we consider a dataset consisting of non-uniform onsite losses and resonant frequencies in a system of $8$ coupled ring resonators. Extending the model to increasingly complex architectures
could potentially require broader training datasets
and retraining over a wider parameter range to further improve robustness and scalability. This provides a natural extension of the present work.

\section{Emulating experimental data from 
theoretical simulations}

\begin{figure}[t!]
\includegraphics[width=\linewidth]{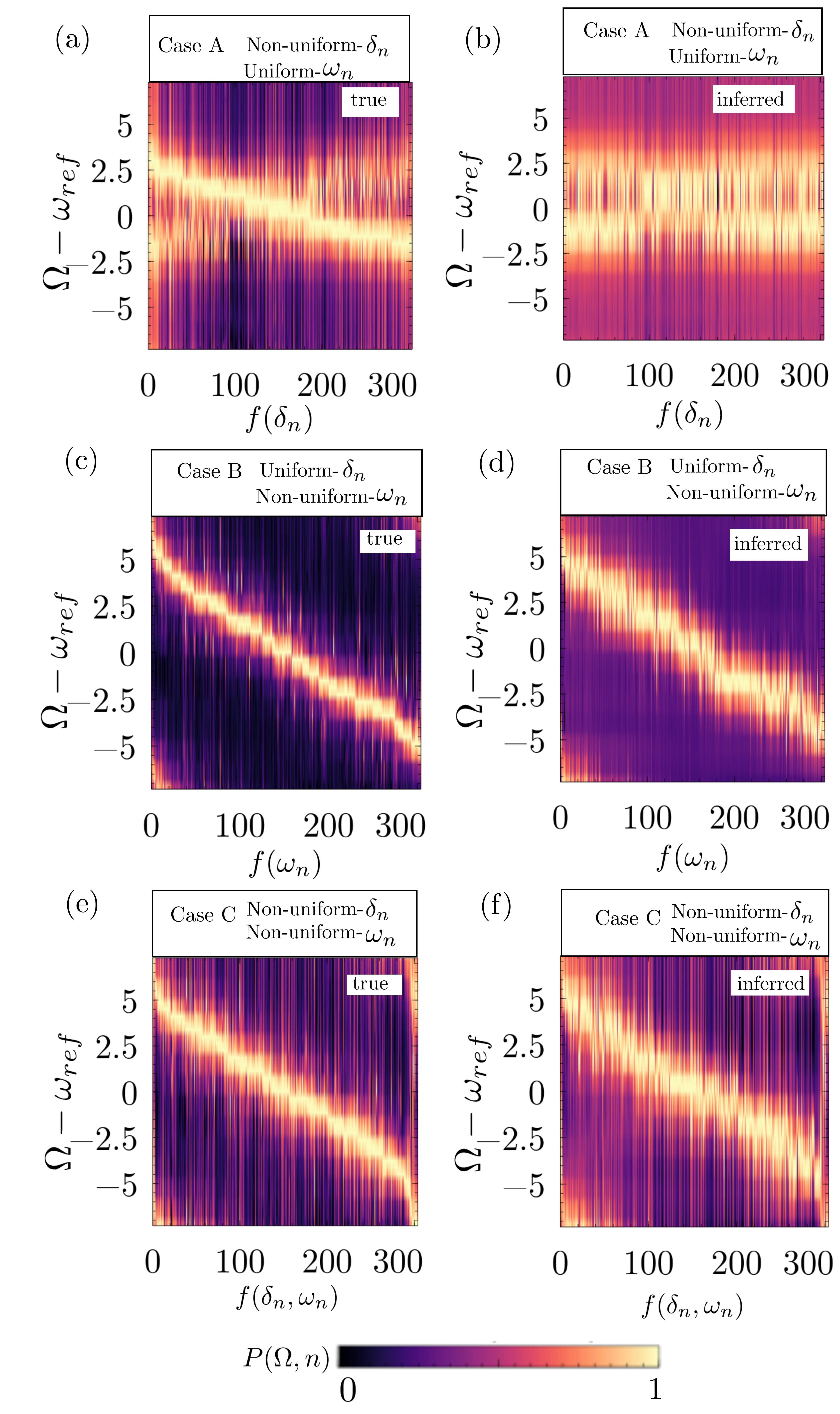}
\caption{(a-f) show normalized spectral power $P(\Omega,n)$ as color. Panels (a–b) show spectra sorted by a function of the onsite losses $f(\delta_n)$, (c–d) by the resonant frequencies $f(\omega_n)$, and (e–f) by the joint metric 
$f(\delta_n,\omega_n)$. In all cases, the ordering ensures that the dominant spectral peak shifts monotonically to higher frequencies, enabling a direct visual comparison between true and learned spectra. The ML-inferred spectra reproduce the correct spectral shape, with most discrepancies limited to the absolute intensity scale.
}
\label{panel:3}
\end{figure}

We further validate our parameter learning approach by reconstructing the experimental spectra from 
theoretical simulations. For this task, we generate the theoretical spectra using the inferred parameters in Eq. \ref{green} and subsequently use a one-dimensional convolutional neural network (1D-CNNs)\cite{Fukushima1980,726791,https://doi.org/10.48550/arxiv.1207.0580} tailored for spectral regression to reconstruct the experimental spectra, as schematically depicted in Fig.~\ref{panel:1}(c).  This model processes each spectrum as a structured signal, allowing the network to capture local spectral correlations and smooth variations that fully connected architectures cannot easily learn. We used regularization based on batch normalization, dropout, and L2 weight decay 
to enhance generalization and reduce overfitting. The network depth and capacity were adjusted across \textit{cases A–C} to balance expressiveness with stability
of the reconstructed experimental spectra from the
theoretical simulations.

In Fig.~\ref{panel:3} we show the comparison between the inferred experimental spectra and the true experimental spectra. We visualize the full training set by reordering the experimental spectra $P(\Omega)$ (normalized for each parameter $\lambda_n$) with an index $f(\lambda_n)$, where $\lambda_n = \delta_n$ in case A and $\lambda_n = \omega_n$ in case B. The mapping $f$ is chosen so that the spectra are arranged in increasing order of their dominant peak frequency $\Omega-\omega_{\mathrm{ref}}$, as shown in Fig.~\ref{panel:3}(a–f). For case C, $f$ depends on both 
$\delta_n$ and $\omega_n$, before sorting by peak frequency, we assign $f$ in increasing order according to the distance from the origin in the $(\delta,\omega)$ plane, so that points closer to the origin receive smaller values of $f$.

Fig.~\ref{panel:3}(a–f) compares the spectral power predicted by the neural network with the experimentally measured spectra for cases A, B, and C. Across all three configurations, the network reliably reproduces the dominant spectral features, with particularly strong accuracy in the positions of the resonant peaks. Small deviations appear mainly in the absolute peak heights, which are more sensitive to experimental noise and amplitude fluctuations than to the underlying physical parameters. The output spectra of the system depend on both the onsite losses and the resonant frequencies. In Case A, varying the loss leads to spectra featuring multiple peaks, which, when they are reordered by the mapping $f$, lead to an
apparent merging and splitting of broad peaks. Since all rings share the same resonant frequencies in this case, the variation in the spectra arises
solely from the different onsite loss configurations. In case B, 
a net evolution of the peaks appears due to the change in the resonance frequency.
The changes in the peak bandwidth originate from the varying configurations of the
resonant frequencies. When neighboring sites have similar frequencies, a large width emerges, whereas when neighboring sites have very different frequencies, very narrow peaks appear.

To quantify the reconstruction accuracy, we compute the residual error, the difference between inferred and true spectral power $|P(\Omega,n)_{\mathrm{inferred}}-P(\Omega,n)_{\mathrm{true}}|$, averaged over all test spectra in each case. The mean residual is on the order of $10^{-3}$
and remains below $3\times 10^{-3}$ in all the three cases. This confirms that the learned model captures the essential spectral structure with good fidelity. Importantly, the inferred spectra remain accurate even when both loss and resonant-frequency variations are present simultaneously (case C), showing that it can extract both $\delta_n$ and $\omega_n$ reliably from the same measurement. 

Reconstructing the full spectral shape, rather than isolated features, suggests that the network captures the noise model and corrections to the model Hamiltonian
present in the experimental measurements.
The agreement between inferred and experimental spectral power supports the one-to-one correspondence between
Hamiltonian parameters and experimental spectra.
This indicates that spectral measurements alone contain sufficient information to infer the effective system Hamiltonian, helping bridge the gap between experiment and theory in integrated photonics.

While these results demonstrate a good accuracy
of the reconstruction, the approach has natural limitations. The model inherits the information content of the spectra; parameters that weakly imprint on the spectral response
cannot be uniquely recovered. In addition, the network is trained on a finite sample of the device configurations, and its performance may degrade for operating regimes beyond the
phase space sampled by our experiments.
Finally, experimental drifts such as fluctuations in laser intensity and frequencies or non-ideal calibration can introduce small systematic errors that the model does not explicitly account for. These constraints define the practical boundaries within which spectral-to-parameter inference remains reliable. The machine-learning model developed in this work provides significant advantages over conventional fitting and optimization approaches, as it is more resilient to substantial noise in the data and does not require assuming a predefined functional form for the spectra.

\section{Conclusion}

We have demonstrated a scalable machine learning strategy to extract intrinsic parameters in coupled photonic systems from global spectral measurements. Using spectral power data from eight coupled ring resonators, we trained neural networks to infer onsite losses and resonant frequency shifts across three experimental configurations. This strategy can enable bypassing the need for device-level calibration or full analytical modeling.
Interestingly, it enables per-ring characterization directly from measured spectra, an ability that is particularly valuable in fabricated platforms where cross-talk and nonuniform losses make traditional tuning challenging. In this sense, the approach provides a direct and data-driven handle on the parameters that govern the behavior of the system, offering a practical tool for tuning, stabilizing, and diagnosing complex programmable photonic circuits.

Looking ahead, this framework can be extended to additional Hamiltonian parameters such as coupling coefficients and programmable phase settings, and scaled to larger arrays where correlations and nonlocal effects become more prominent. Incorporating temporal or broadband measurements would allow the model to capture richer multimode dynamics. Finally, embedding the inference model in a feedback loop opens a route toward closed-loop optimization and autonomous control of reconfigurable photonic architectures.

\section*{Data Availability Statement}
The data that support the findings of this study are openly available in \cite{elizen}.
\section{Machine Learning Architecture}
\subsection{Regression task - learning of intrinsic parameters}
Here we describe the neural-network architecture used to infer the onsite losses $\delta_{n}$ and resonant frequency shifts $\omega_{n}$ from the measured spectral power distributions in detail.

For the inference of intrinsic losses in case A, we collected output spectral power for $200$ eight-ring systems ($1600$ samples). The fully connected network was trained on $1280$ samples and validated on $320$, using $40$ features. Training was carried out for $2
00$ epochs with a batch size of $16$. All layers used ReLU activations\cite{10.5555/3104322.3104425,pmlr-v15-glorot11a}, Glorot uniform weight initialization and Adam optimizer\cite{Kingma2014AdamAM,j.2018on} at a fixed learning rate of $0.001$. The neural network architecture for Case A is such that it has a total number of input parameters of size $40$ (i.e. the size of features of experimental spectral density), with four dense layers with nodes $40,20,10,$ and $1$, respectively, to give a total of $2681$ trainable parameters. The mean squared error served as the loss, and the inputs and targets were independently standardized prior to training.

For the regression task that infers resonant frequency shifts as in case B, spectra were produced for $300$ eight-ring systems ($2400$ samples) and reduced to $1631$ after filtering\footnote[1]{Since a photonic integrated circuit inherently contains a resonator featuring multiple modes, its measured spectra naturally reflect this multimodality. However, as our model is based on a tight-binding description with unimodal sites, we restrict the analysis to the spectral density within the frequency range of $(-37.7 + \omega_{\mathrm{ref}}, 37.7 + \omega_{\mathrm{ref}})$ to isolate the relevant single-mode response.}.
Of these, $1304$ was used for training and $327$ for testing with the architecture. The hyperparameters matched the previous experiment, except that the learning rate used a learning rate scheduler. Notably, the neural network architecture for Case B is such that it has a total number of input parameters of size $40$, with six dense layers with nodes $40,30,20,10,5$ and $1$, respectively, to give a total of $3761$ trainable parameters. 

Finally, the joint extraction of frequency shifts and losses (case C) used the same filtered dataset ($1304$ / $327$ split) and a dense network, again trained for $200$ epochs with batch size $16$, ReLU activations, Glorot initialization, Adam with the scheduled learning rate and mean squared error loss. For this case, we use a wide fully-connected neural network, it has a total number of input parameters of size $40$, with four dense layers with nodes $800,600,400$ and $2$, respectively, to give a total of $754602$ trainable parameters. The number of output nodes is $2$ to learn both loss and resonant frequencies.

\subsection{Inverse problem - mapping theoretical spectral densities to experimental }

Here we describe the neural-network architecture used to map the theoretical spectral power distributions computed from the inferred parameters to their experimental counterparts.

For the inference of the experimental spectra in case A, we use all the parameters similar to those for the learning of onsite losses, however the neural network architecture is different here, the output nodes are $40$. We employed a convolutional neural network (CNN) to emulate experimental data from theoretical simulations rather than a fully connected neural network architecture, as the CNN exhibits reduced overfitting and lower mean squared error as the number of training epochs increases. We use a regularized 1-D CNN\cite{Fukushima1980,726791,https://doi.org/10.48550/arxiv.1207.0580} for spectral regression. The model adds Gaussian noise to the input ($\sigma=0.02$), then applies three Conv1D layers ($32, 64, 64$ filters with kernel sizes $7, 5,$ and $5$) with ReLU, batch normalization, L2 weight decay ($5\times 10^{-4}$), and dropout ($0.2$), including a single two times max-pool after the second convolution. Features are aggregated by global average pooling and passed to a $128$-unit ReLU dense layer (dropout $0.4$) and a linear output whose length equals the input length. 

For the inference of experimental spectra, the CNN of case B follows case A but is lighter, two 1-D conv layers ($32$ and $48$ filters with kernel sizes $7$ and $5$) with batch-norm and ReLU. We apply spatial dropout ($0.15$) after the first conv, downsample with a two times max-pool, use global average pooling, then a small $64$-unit dense layer (L2$=1\time 10^{-4}$, dropout $0.25$), and a linear head that outputs a length-L spectrum ($40$ by default). 

Also, case C follows the same 1-D CNN pattern as cases A/B but is larger, with three Conv1D layers ($64, 128, 128$ filters with kernel sizes $7, 5, 5$) with ReLU and batch-norm, dropout $0.2$ after the first two convs, a two times max-pool after the second conv, and global average pooling. A $256$-unit ReLU dense layer with L2$=1\times 10^{-3}$ and dropout $0.4$ feeds a linear output of length L (equal to the input length).

\section{Data collection}
We collected the experimental data using a programmable integrated photonics platform described in Ref.\cite{On2024,amin2024}. The setup consists of a tunable laser source, photodetectors, and a programmable photonic chip based on a two-dimensional hexagonal mesh. Each arm of the hexagonal ring contains a programmable unit cell (PUC), implemented as a Mach–Zehnder interferometer with thermo-optic phase shifters that control both the power splitting and the relative phase, enabling full programmability of the circuit.

This platform has previously been used to realize the Hermitian Su–Schrieffer–Heeger (SSH) model by tuning the couplings between adjacent rings through the PUC phase settings \cite{On2024}. It also supports non-Hermitian physics through site-dependent loss engineering, where each onsite loss $\delta_n$ is controlled by the PUCs on ring $n$. Using this capability, the non-Hermitian Aubry–André–Harper (AAH) model \cite{ourprr} was recently implemented by modulating the onsite losses \cite{amin2024}, with all rings set to the same resonance $\omega_{\mathrm{ref}}$.

To study a more general coupled-ring Hamiltonian, we extended this scheme by introducing individual detunings $\omega_n-\omega_{\mathrm{ref}}$ through the phase of one PUC in ring $n$th, and by assigning different losses $\delta_n$. Each 
$\delta_n$ consists of the intrinsic ring loss combined with a tunable loss set by coupling the ring to an output port. These losses can be independently determined using the add-drop filter calibration method described in the Supplementary Information of \cite{amin2024}. In our experiments, each parameter pair $(\delta_n,\omega_n)$ with $n\in [0,7]$ therefore specifies a distinct programmable photonic circuit comprising an array of eight coupled optical resonators.

The internal laser of the PIC is centered at the resonant wavelength of $1550.779$nm and can be swept with a wavelength resolution of $3$pm, allowing us to record the spectral power response of the array under a wide range of programmed configurations.

\textbf{Acknowledgments}
We acknowledge the financial support of the 
Nokia Industrial Doctoral School in Quantum Technology.
JLL acknowledges the 
financial support from the
Research Council of Finland
Project No. 370912,
the Finnish Quantum Flagship,
InstituteQ,
the Finnish Centre of Excellence in Quantum Materials (No. 374166),
and ERC CoG ULTRATWISTROICS (no. 101170477).
AB-R acknowledges support by the National
Science Foundation (NSF) (award ID 2328993).
ELP, FA and JLL acknowledge the computational resources provided by
the Aalto Science-IT project.

\bibliography{biblio}

 \end{document}